# The Creation of the World – According to Science

**Ram Brustein, Judy Kupferman**

Department of Physics, Ben-Gurion University, Beer-Sheva 84105, Israel
CAS, Ludwig-Maximilians-Universitat Muenchen, 80333 Muenchen, Germany
E-mail: ramyb@bgu.ac.il, judithku@bgu.ac.il

**Abstract**

How was the world created? People have asked this ever since they could ask anything, and answers have come from all sides: from religion, tradition, philosophy, mysticism… and science. While this does not seem like a problem amenable to scientific measurement, it has led scientists to come up with fascinating ideas and observations: the Big Bang, the concept of inflation, the fact that most of the world is made up of dark matter and dark energy which we can not perceive, and more.

Of course scientists cannot claim to know the definitive truth. But we can approach the question from a scientific viewpoint and see what we find out. How do we do that? First, we look to the data. Thanks to modern technology, we have much more information than did people of previous ages who asked the same question. Then we can use scientific methods and techniques to analyze the data, organize them in a coherent way and try and extract an answer. This process and its main findings will be described in the article.

# Introduction

How was the world created? People have asked this ever since they could ask anything, and answers have come from all sides: from religion, tradition, philosophy, mysticism….and science. While this does not seem like a problem amenable to scientific measurement, it has led scientists to come up with fascinating ideas and observations: the Big Bang, the concept of cosmic inflation, the fact that most of the world is made up of dark matter and dark energy which we can not perceive, the fact that in every direction we observe the same very faint background radiation, and more.

 Of course scientists cannot claim to know the definitive truth. But we can approach the question from a scientific viewpoint and see what we find out. How do we do that? First, we look to the data. Thanks to modern technology, we have much more information than did people of previous ages who asked the same question. Then we can use scientific methods and techniques to analyze the data, organize them in a coherent way and try and extract an answer.

The concept of creation takes on a particular and specific meaning in a scientific context, not to be confused with the concept of "creation out of nothing" that we find in metaphysics or in monotheist theologies. In its narrow and most commonly used sense, it means a specification of the state of the universe at some initial time, together with the laws of physics that have evolved this initial state up until today. The initial state may or may not be approximately classical or quantum and the laws of evolution may involve quantum mechanical equations or classical equations. Sometimes the specification of the initial state is only statistical, chosen from some ensemble of states with a prescribed probability. In this case, the idea of one initial state is replaced by the set of possible initial states and the probability distribution on it. Even when Stephen Hawking describes the creation of the Universe from "nothing" the process involves a specification of some initial conditions for the quantum wavefunction. So in order to discuss creation, we need to consider what may have been the initial conditions. Thus, the scientific meaning of "creation" is in effect a mathematical description in terms of equations and initial conditions of a "natural beginning" or an "emergence from something".

## The universe today

Since we wish to know whether the universe had a beginning and if so, how the universe began, it would help to construct a picture of the early universe – what was it like at the earliest possible times? We do this by looking at the universe today. We know a lot about the laws of nature today, and we have many indications that they have not changed in the course of the universe's lifetime. So we can use them to try and construct a picture of the early universe. We can look at the universe today – its content and its size and its development – and try to extrapolate backward. Another complementary way of learning about the state of the universe at early times relies on Einstein's theory of special relativity. This theory says that light from far away had to travel a long time. So the light we observe today from distant sources was emitted when the universe was much younger, and provides information about a time long ago.

When we look at the world today, what do we find? We begin with what we can see. It turns out that we can't see much! Very little of the universe is actually visible matter, in fact only about five percent. This is made up of stars and gas (mostly hydrogen), all bound together by gravity into galaxies. The galaxies too are bound together, organized into clusters.

> **Length Scales in the Universe**
>
> A useful unit of distance is the parsec, which is the characteristic distance between stars.
>
> - 1pc=3.26 light years – about 30 billion kilometers.
> - Typical galaxy size: 10 kiloparsec, or 30,000 light years.
> - Distance between galaxies: 500 kpc, or about 1.5 million light years.
> - Distance to the galaxy cluster nearest us: 20 Mpc (million parsecs)
> - Size of the visible universe : 10Gpc (a gigaparsec is a billion parsecs), about 30 billion light years.

Stars are spherical bodies made up mostly of hydrogen. A star emits light because it has a natural nuclear reactor inside, burning "on a low flame". There are about a hundred billion stars in a galaxy, and a hundred billion galaxies in the visible universe – that is altogether $10^{22}$ stars (10,000,000,000,000,000,000,000). The galaxies turn round and round, at the breathtaking speed of one complete rotation every hundred million years.

> In fact there are far more stars than grains of sand on the shore! We can work this out:
> - The average size of a grain of sand is 1 mm. so there are a billion grains of sand per square meter.
> - In one kilometer of sea shore there are about ten thousand square meters – that is about $10^{13}$ grains of sand.
> - Israel has a thousand km. of seashore – $10^{16}$ grains of sand! That is six orders of magnitude (a million times) smaller than the amount of stars in the sky.

**What else does the universe contain?**

If visible matter is only about 5% of the universe, what else is there? About a quarter of it is invisible, and is therefore called "dark matter," within and surrounding the galaxies and the clusters. There is about six times more dark matter than visible matter! But how do we know it is there? Dark matter exerts the force of gravity on visible matter. We can see this in two ways. First, we measure the speed of rotation of stars and estimate from the velocity the strength of the force that is driving the rotation and from that the amount of matter that is exerting this force. Second, we "look" at galaxy clusters.

An example of a famous galaxy cluster is the Perseus cluster. How can we map out the dark matter in a galaxy cluster? By charting the proper velocities of individual galaxies and stars, by looking at the temperature map, by analyzing gravitational lensing and by reconstructing collisions. We conclude that in galaxy clusters, too, there is about five times as much dark matter as visible matter.

So far we have about 5% visible matter, and then another quarter which is dark matter – that leaves a large chunk of unidentified stuff. We call the remaining constituent of the universe "dark energy", and it is spread uniformly throughout the entire universe. How do we know? That is a long and fascinating story, and it is not yet complete. That story should be told in another article and we will not attempt to tell it here.

**How does the universe behave?**

Now we have looked at the universe and described what it contains. The next question is: what is it doing? Most people have heard that it is expanding. People often ask: expanding into what? One popular explanation is that the universe is a sort of balloon. We draw stars on the surface of the balloon, and as we blow it up, we see the stars going farther apart. But the balloon expands into the surrounding air. The universe, however, has no surrounding air. It's all there is. So into what does it expand? The correct answer is – into nothing. There is nobody outside the universe watching it grow bigger and bigger, as you might watch the balloon. Instead, the expansion can be understood as a recalibration of distance. This was Albert Einstein's major discovery in 1907 that led to the general theory of relativity, completed 10 years later.

Picture a drawing of a grid. Say the grid lines are a centimeter apart. Now draw two stars, each on a grid line and with one grid line between them. So the stars are about two centimeters apart. Now somebody waves a magic wand, and the grid lines change slowly until they are now a meter apart. The stars are still sitting on the same grid lines. They haven't moved with relation to the grid, and they haven't moved outwards into some outer space. But they are now a hundred times further apart, just because the measure of distance between them has grown.

**How do we know it's expanding?**[1]

---

[1] An account of the history of the discovery of the expanding universe can be found, for example, in Harry Nussbaumer and Lydia Bieri, arXiv:1107.2281v2 [physics.hist-ph] and in Marcia Bartusiak, "The Day We Found the Universe," Pantheon Books, 2009.

Galaxies emit light in different colors. The redder the light, the longer its wavelength and the lower its frequency. On the other hand blue light has a shorter wavelength and higher frequency. We find that emission lines from gas from far away galaxies are shifted to the red end of the frequency.

Hubble's law, discovered by Edwin Hubble in 1929, tells us that the further away the light emitting object is from us, the greater its red shift. The law relates a "fake velocity" and distance by a formula: $cz = H_0 d$, where c is the speed of light and z the red shift, so that cz together gives the "fake velocity". The velocity is a fake because it is not the galaxies themselves which are moving, just as the stars on the grid above are not moving but rather the grid is expanding. The Hubble constant $H_0$ is a constant of proportionality, with units of 1/second, and d is the distance. The formula means that the red shift is proportional to the distance: the further away the light emitting object is, the redder it appears. In this way we can tell as galaxies look redder that in fact they are going farther away.

> **The discovery of the expanding universe**
>
> The Russian Alexander Friedmann was the first to discover time-dependent cosmological solutions to the Einstein equations and to understand that in some of them the universe is created at some instant of time in the past. In his first 1922 paper he actually calculated the age of the universe since its creation and found that it is about 10 billion years, a surprisingly accurate number. It is clear that Friedmann understood the relationship between the age of the universe and its expansion rate. If one translates the age of 10 billion years into an expansion rate one gets a number which is much closer to the correct value than the number that Lemaître and Hubble later obtained (see below).
>
> In 1927 the Belgian priest and cosmologist Georges Lemaître, while looking for a way to combine the static matter-filled universe of Einstein with the empty expanding universe of the Dutch astronomer Willem deSitter, independently rediscovered Friedmann's solutions, and for a particular model he was able to use the redshifts and distances of nebulae known then to obtain the relation that would later become known as the "Hubble law". Lemaître along with George Gamow emphasized the concept of "natural beginning" of the universe.
>
> It is sometimes argued that Friedmann and Lemaître receive less credit for the discovery of the expanding universe due to "sociological reasons", that they were not as well known as more famous scientists such as Sir Arthur Eddington, Einstein or deSitter, or because their original work is written in less familiar languages. Without going into the details of this debate, let us just say that in our opinion this argument is inadequate because the scientific work of both was well known to the leading cosmologists. The simpler and better explanation is that the significant contributions of Friedmann and Lemaître were not the central contributions to the main thrust of developing the idea of the expanding universe.

We set out to look at the universe today as a basis to asking about its beginning. What do we know? We have seen what the universe contains: 5% of visible matter, another 1/4 dark matter and the remainder is something we don't know, but which we call dark energy. We also know that it is expanding. And we know quite a bit about the visible matter. Based on what we know about the universe, scientists have more than one suggestion as to how it began.

**The Hot Big Bang**

The Hot Big Bang model of the universe proposes that at earlier times the universe was hot and dense. As we look back in time we see two substantial changes: First, expansion thins things out. As the universe expands, since new matter is not created, the density of matter becomes smaller. So the density of matter at early times was greater.
Second: As it expands the universe is cooling off. The temperature is a measure of the average velocity of particles. Now imagine two particles (they could be gas molecules or even entire galaxies) that are no longer at rest but rather move at a certain speed. Since the grid is expanding, they cover fewer grid points at the same time than if there were no expansion. This means that their velocity is decreasing and therefore so is their temperature. So the universe was once hotter.

What proof is there of the Hot Big Bang model? There are three major pieces of evidence. The first, which we have just discussed, is the expansion of the universe. Another significant indication is the existence of faint uniform radiation wherever we look. This is called cosmic background radiation and has led to two Nobel prizes: in 1978 to astronomers Arno Penzias and Robert Wilson who discovered it, and in 2006 to John C. Mather and George F. Smoot, who analyzed observations of the radiation and found that it confirms many aspects of the Big Bang theory.[2] The third piece of evidence relates to the creation of the elements: nucleosynthesis.

**Cosmic background radiation**

Everywhere astronomers look they detect a uniform general background of radiation. This background radiation is a remnant of times when the universe was much hotter. Mather and Smoot's analysis of data from the COBE satellite showed that the radiation has a black body spectrum, that is, a spectrum dependent only on temperature, and which today is barely 2.7 degrees above absolute zero. This fits the picture of the early universe as a glowing body which has cooled off. In addition they found tiny relative variations of temperature from place to place of about 1/100,000 of the average temperature. These variations give indications as to how galaxies and clusters of galaxies began to form from an almost uniform universe.

The Big Bang model asserts that the universe was hotter in the past, so the radiation itself had to be hotter in the past. Recently, it has actually become possible to verify that radiation was hotter in the past! At earlier times, the radiation was hot enough to excite carbon atoms in ways that colder radiation cannot. The excited atoms are illuminated by light from a distance strong source and absorb it at a characteristic frequency, thus giving rise to particular absorption lines in the observed light. Once telescopes became powerful enough, these lines were detected, providing the long sought after proof.

---

[2] See, for example, http://www.nasa.gov/vision/universe/starsgalaxies/nobel_prize_mather.html

## Creation of the elements (nucleosynthesis)

When the temperature of the universe was 10 billion degrees it contained a hot soup of neutrons, protons, electrons and positrons, light (photons) and neutrinos. It cooled off for about three minutes and then hydrogen began to form, then "heavy water" (deuterium) and after that helium as well and a very small amount of lithium. This process is called "Big Bang Nucleosynthesis." It was first discussed in a paper by Ralph Alpher, Hans Bethe and George Gamow in 1948[3] and later improved and refined. Simple considerations allowed them to estimate the relative ratio of helium to hydrogen. Since hydrogen has one proton and helium has two protons and two neutrons, the ratio of their densities is determined by the ratio of number of neutrons to protons at the time that helium could be created. Putting in the known properties of protons and neutrons yields the prediction of the Big Bang theory: 25% helium. The prediction is verified to a large degree of accuracy!

All heavier elements, which include a larger number of protons and neutrons than helium, could not have been created from the cosmic soup because its density and temperature were by then too low to facilitate their creation. So they must have been created later by nuclear fusion out of lighter elements in the cores of stars such as our sun, where the temperatures and densities are high enough. All visible matter in the universe is made of this stuff, not just stars. So everything that we see around us, earth and rocks and animals and even we ourselves are made of stardust!

## Reconstruction of the early universe in accelerators

Another way to get an idea of the early universe is to try and determine the laws of physics that were relevant to the evolution of the universe at early times and even try to recreate the conditions that we believe existed then, and see what happens. Accelerators are huge machines which can smash a few hundreds particles together at enormous speeds and allow us to realize this dream, at least partially. A more detailed description of this vast topic deserves a much expanded discussion which we will not attempt here. The interested reader can consult several excellent books on the subject.[4]

## Inflation

The Hot Big Bang model asserts that the universe was once hot, dense and smooth. From this assumption by using the known laws of physics we can reconstruct its development into the universe we see today. But there are some intriguing questions. First, why was the primordial universe so smooth? In fact it seems to be too smooth, to the degree that points in space that are too far from each other to have been in causal contact have the same temperature. Second, why is it so old? And third, why is it hot?

---

[3] R. A. Alpher, H. Bethe, G. Gamow, "The Origin of Chemical Elements," Phys. Rev. 73, 803 (1948).
[4] For example, B. Greene, "The Elegant Universe: Superstrings, Hidden Dimensions, and the Quest for the Ultimate Theory", Random House, 2000.

The accepted paradigm for explaining the initial state for the Hot Big Bang model of the universe is cosmic inflation. The idea is that the very early universe has undergone a rather long period of accelerated expansion making its final radius larger by a factor of about $e^{60} \sim 10^{25}$ from the initial radius. The idea of inflation was expressed most clearly by Alan Guth in 1982.[5] From Einstein's equations we know that to enter such a phase of accelerated expansion, the universe had to be filled with some constant and high energy density during this epoch. We know that the late universe is undergoing a phase of accelerated expansion (recall the discussion of dark energy) so such epochs are physically possible.

The accelerated expansion has several effects. First, the effect of smoothing things out. Imagine a small perturbation of a flat universe. For example, it could be a blob of slightly denser radiation. Now when the universe expands in an accelerated way its volume increases exponentially so the density of matter decrease exponentially and differences in the matter density also decrease exponentially. So the expansion itself acts a bit like an iron, smoothing out a piece of cloth till it lies flat from one end of the ironing board to the other. The second effect is to allow points which today are too far apart in space to have causal interactions between them to have been in causal contact in the past. For instance take two points on the grid and the blob of slightly denser radiation that extends through all the area between the two points that we mentioned above. As universe expands, and these points grow farther apart the blob still extends from one point to another, but meantime it goes through a much larger area of space than it did before. If the expansion at one time was accelerated, the two ends of the blob will seem to be too far from each other to allow light to propagate from one end to the other.

Acceleration ages: a spherical universe of typically small size would tend to collapse on itself in a small amount of time. If it underwent a long period of inflation, its size would exponentially increase and so would the time that it would take it to collapse.

Acceleration heats: After the end of the era of inflation, the energy of expansion is transformed into hot matter. Thus all the matter in the universe was created, as well as its structure.

Acceleration hides the past: Accelerated expansion creates a causal barrier – a horizon between the future (today) and the past eras before inflation started. An observer in the future sees only a very uniform ball of fire with a temperature that decreases with time. The slight fluctuations in temperature in this ball of fire originate from quantum fluctuations during inflation. These tiny perturbations constitute the seeds that have been amplified by gravity and grown into the galaxies and cluster of galaxies that we observe in the universe.

Can we prove inflation? This is hard and perhaps impossible. Inflation is a paradigm. To be able to prove or disprove inflation, we need specific predictions that can be tested by experiment and we need to verify that these predictions cannot be a result of a different theory. The generic predictions of inflation are already verified by experiment, a spatially flat universe and a specific spectrum of primordial cosmic perturbations. But will we say that inflation is incorrect if it is found by future observations that the spatial curvature of the universe is small but nonvanishing? Or if the spectrum is found to be not exactly flat? The answer is no, because there are models of inflation which do make such predictions. Specific predictions are obtained from specific models of inflation. Those are complicated and sometimes have overlapping predictions, so even if one is disqualified as a description of nature the others still survive.

---

[5] A. Guth, "The Inflationary Universe," Perseus Publishing, 1998.

Another complication is that inflation does not have a real competitor theory that can make predictions for all aspects of cosmology that inflation can. All the competitors of this type which existed fell by the wayside as measurements became more accurate. On the other hand, there are several alternatives for each specific aspect that inflation predicts.

**The Big Bang initial singularity/explosion**

The solutions to Einstein's theory of general relativity have the property that every model of the universe shows that looking backwards; it reaches a point where the equations no longer hold. We call this era the "initial singularity." It is sometimes referred to as the "big bang" singularity or simply the "big bang." The term "big bang" is meant to create an image of a big explosion that started at a point. However, this image is misleading. The correct concept of the big bang singularity should be that of an explosion that occurred simultaneously at each spatial point in the universe.
When it exploded, the universe itself could have been very large or even infinite. It does not necessarily shrink to a single point; rather what typically occurs is that its rate of expansion or contraction can become so large that the Einstein equations loose their validity. Another possibility is that the universe becomes so anisotropic that the Einstein equations can no longer describe it. In technical terms, the equations that govern its evolution break because the curvature of the universe becomes formally infinite.

If the era of the initial singularity is followed by an era of cosmic inflation it becomes hidden from us as future observers by the horizon created by the accelerated expansion. So this phase will be very difficult to probe or even to show that it actually existed. This has not stopped theoretical physicists from speculating about its properties. The lack of data may even have encouraged them to be wilder in their speculations…

There are a few different ideas about this. The first is due to Stephen Hawking.

**Quantum universe:**

In quantum theory, the probability that a particle winds up at a certain spot is calculated by summing up all its possible paths. The particle actually goes through all the possible paths at once. Stephen Hawking claimed in the 1980's that this is true of the universe as a whole: it too must evolve through many simultaneous histories. The world we see today is a sum over all these histories.[6] That is, all the histories happened, but some of them cancelled out and others added up together, and the universe that we see is the superposition of all the histories that did not cancel out.

**Pre Big Bang**

The standard Big Bang theory has it that at the beginning the distance between everything was zero, and before that there was nothing. Time itself has no meaning as a concept before that. More sophisticated models which take quantum effects into account and use elements from string theory argue that things must have begun at a certain distance apart. These models lead to the possibility of a universe before the big bang. In such models too there was a big bang, but it was not the beginning of everything but only a transition, resembling a strong explosion.[7]

In the distant past, according to according to a model proposed by Gabriele Veneziano in 1991, the universe was nearly empty, and forces such as gravity were very weak.[8] They gradually strengthened and matter began to clump. At some points it clumped so densely that a black hole was formed. Inside the hold matter fell to the middle and increased in density to a maximum possible density, and then quantum effects caused it to rebound into a big bang. Outside the black hole, where matter was completely cut off from the matter inside, other holes began to form – each of them into a separate universe.

If the phase of the big bang was not followed by cosmic inflation then a distinct signature of the explosion can be observed even today: a background of gravitational radiation similar to the background of electromagnetic microwave radiation[9].

**Ekpyrotic universe**

Another model is called the Ekpyrotic model of the universe.[10] Ekpyrosys means a sudden burst of flame in Greek. This model proposes that the universe at its beginning was not hot and dense, but rather cold and nearly empty. Then there was a collision, the "sudden burst of flame," as a result of which it became hot and began to expand. This was a collision of two different 3-dimensional worlds moving in a space with an extra 4$^{th}$ spatial dimension. The kinetic energy in the collision was converted into electrons, photons and other elementary particles, which were confined to three dimensions. The temperature after the collision was finite, so that there was no singularity in fact. This model is based on currently unproven ideas from string theory and has several conceptual and technical problems.

---

[6] S. Hawking, Black Holes and Baby Universes and Other Essays, Bantam Books, 1993
[7] G. Veneziano, The Myth of the Beginning of Time, Scientific American, May 2004.
[8] Although the laws of nature have not changed in the course of the universe's lifetime, the coupling parameters – the strength with which forces act – may have done so.
[9] R. Brustein et al., Relic gravitational waves from string cosmology, Physics Letters B361 (1995) 45.
[10] P. J. Steinhardt and N. Turok, "Endless Universe: Beyond the Big Bang," Doubleday, 2007.

**Other explanations**

The above are some scientific explanations for the creation of the world. But there are explanations outside of science as well. One interesting possibility is that – there is simply no explanation. We can't explain it because there was no definite reason that it happened. It just did. This is not as outlandish as it sounds. It is perfectly reasonable to think that not everything has a reason! Reasonable people don't enjoy thinking so – but it is still a possibility. Maybe something happens just because it does.

Another possibility, called the anthropic principle, is that the universe is what it is because this is the result most suited to human life. As pointed out by Robert Dicke in 1961, the age of the universe as we see it cannot be random; if it were older or younger we would not be here to see it. [11] The term "anthropic principle" was coined in 1973 by Brandon Carter, and he formulated it in two versions: The weak anthropic principle holds that physical and cosmological facts are not all equally probable, but that they take on the specific values that we observe because only those values lead to a world where life is possible. The strong anthropic principle holds that the world must be such as to lead to the existence of observers in it.[12] This may sound something like certain religious ideas, but the viewpoint is scientific: it is based on values of observed physical quantities, in conjunction with a certain viewpoint in quantum mechanics which holds that the collapse of a wave function into an observable value is due to interaction with an observer.

Critics have pointed out that since it is not a falsifiable idea, it does not really belong to science. Another criticism is that the anthropic "principle" is not really a scientific principle. A scientific principle could be defined as a general law from which specific laws of nature in the form of mathematical equations can be derived in specific circumstances. A famous example in physics is the Heisenberg uncertainty principle. From this point if view the use of the word "principle" in this context is misleading, as we explain below. Perhaps a more appropriate term would be "anthropic conditions".

**Life supporting universe?**

The idea on which the anthropic principle is based is that it is possible to constrain theories and models of the universe by the aposteriori requirement that the conditions for the existence of "life" are obeyed. This idea has some fundamental difficulties. First, it requires a working definition of "life". As a scientific subject this is a very complicated issue.[13] So far, it is unclear which of the ingredients and parameters are essential for "life". In most analysis in physics "life" is replaced with a much simpler condition that is argued to be a necessary condition for the form of life that we know and without any consideration to the possible existence of other forms of life.

---

[11] R. H. Dicke, "Dirac's Cosmology and Mach's Principle". *Nature* **192**: 440–441 (1961).
[12] B. Carter, "Large Number Coincidences and the Anthropic Principle in Cosmology," *IAU Symposium 63: Confrontation of Cosmological Theories with Observational Data*: 291-298, Dordrecht: Reidel (1974).
[13] R. Popa, "Between Necessity and Probability: Searching for the Definition and Origin of Life", Springer, (2004).

The idea of constraining theories by the aposteriori requirement that the universe they lead to support life is useful only in cases in which "life" is highly improbable, which means that from most of the parameter space "life" cannot form. The idea that life is improbable requires, in addition to a definition of life, some idea about the probability that any form of life will arise, which of course is a highly complex issue. Even in cases for which "life" is a possibility, it is possible in a statistical sense, so the process of the formation of life is statistical. This means that for the same values of the parameters life will form in some cases, while in other cases it will not.

Then, of course, there are explanations given by thinkers in other areas of human thought: religion, philosophy, mysticism. The unique contribution of science is not in proposing a definitive answer, but in its ability to investigate the question using scientific methods and tools. Guesses and ideas about the creation of the world can be checked by experiments: by astronomical observations and by recreation, for example, in accelerator experiments. Science grants a real possibility of approaching an answer, but we still have no idea if an absolute answer can ever be found.